\documentclass[12pt]{article}
\usepackage{graphics}
\usepackage{amssymb}
\usepackage{color}
\usepackage{amsmath}
\usepackage{amsthm}
\usepackage{amsopn}
\def\uno{\mbox{1 \kern-.59em {\rm l}}}
\def\nn{\nonumber}
\def\be{\begin{equation}}
\def\ee{\end{equation}}
\def\bea{\begin{eqnarray}}
\def\eea{\end{eqnarray}}

\begin{document}
\begin{center}
{\bf{\large Quantum Correlations in Neutrino Oscillation:
Coherence and Entanglement
}}
\vskip 4em
{\bf M. M. Ettefaghi} \footnote{ mettefaghi@qom.ac.ir  }
 \vskip 1em
 Department of Physics, University of Qom, Qom 371614-6611,
Iran.
\vskip 2em
{\bf Z. S. Tabatabaei Lotfi}, and {\bf R. Ramezani Arani} \footnote{ ramarani@kashanu.ac.ir }	
\vskip 1em
University of Kashan, Km 6 Ravand Road, Kashan 87317-51167, IRAN

 \end{center}
 
  \vspace*{1.9cm}
\begin{abstract}
 In this paper, we consider the quantum correlations, coherence and entanglement, in neutrino oscillation. We find that the $l_{1}$-norm as a coherence measure is equal to sum of the three possible concurrences for measuring the entanglement among different flavor modes which were calculated in the paper by (M. Blasone et al., Europhys. Lett., {\bf 112}, 20007). Our result shows that the origin of the flavor entanglement in neutrino oscillation is the same as that of quantum coherence. Furthermore, in the wave packet framework, the variation of $l_{1}$-norm is investigated by varying the wave packet width $\sigma_{x}$. As it is expected the amount of coherence increases by $\sigma_{x}$ due to the increase in overlapping of the mass eigenstates.
\end{abstract}
\newpage
\newpage
\section{Introduction}

 Quantum resource theories provide a very powerful tool for studying quantum phenomena such as entanglement and quantum coherence \cite{quantum resource theory}. One of the most useful aspects of a quantum resource theory is that it introduces some meaningful and applicable methods to quantify the desired quantum resource and provides us with measures in order to determine the quantumness. The basic operation of quantum resource theory is that it divides all quantum states into two groups: the free states and the resource states. Also, we should define a set of free quantum operators that are natural constraints applied to our set of free states. More explicitly, after the action of these free operators on the set of free states we will again have  a free state. Among the various quantum resource theories, we can name the two famous ones, entanglement \cite{plenio1996} and quantum coherence \cite{plenio2014}. Although there is much freedom in defining free states and free operators, unexpected similarities appear among different quantum resource theories according to the resource measures as well as the resources convertibility. In particular, entanglement and quantum coherence, which seem to be distinct resource theories, show deep constructional and conceptual analogies \cite{plenio2017}. In one of the first approaches in this direction, it is shown that the entanglement measure can be used for measuring quantum coherence \cite{relation between coherence and entanglement, coherence concurrence}, because we can get an entangled state from a coherent and an incoherent state via an incoherent operator. 
 
Investigating quantum effects in neutrino oscillation can be a very interesting issue, since in this case, quantum correlations such as quantum coherence, which is a microscopic quantum effect can be studied in large distances even up to several hundred kilometers away. The idea of neutrino oscillation was first put forward more than half a century ago and since then many experimental evidences showing the transition between different neutrino flavors have been obtained from different sources \cite{historical}. Neutrino and anti-neutrinos can be produced through weak interactions in three different flavor eigenstates, namely electron e, muon $\mu$ and tau $\tau$. Neutrino oscillation involves mixing of flavor states $\vert\nu_{e}\rangle$, $\vert\nu_{\mu}\rangle$, $\vert \nu_{\tau}\rangle$ as a linear superposition of mass eigenstates, $\vert \nu_{1}\rangle$, $\vert\nu_{2}\rangle$  and $\vert\nu_{3}\rangle$ in which(e, $\mu,\tau$) represent flavor states and (1, 2, 3) represent mass states:
\be
\vert\nu_{\alpha}\rangle= \sum_{i} U^{\ast}_{\alpha i} \vert\nu_{i}\rangle. \label{flavor states}
\ee
The unitary matrix U is characterized by three mixing angles $\theta_{12}, \theta_{23}, \theta_{13}$, charge conjugation and the parity (CP) violating phase $\delta_{CP}$:
\be
\left(\begin{array}{ccc}
c_{12}\ c_{13} & s_{12}\ c_{13} &s_{13}\ e^{-i\delta_{CP}} \\ 
-s_{12}\ c_{23}-c_{12}\ s_{13}\ s_{23}\ e^{i\delta_{CP}} & c_{12}\ c_{23}-s_{12}\ s_{13}\ s_{23}\ e^{i\delta_{CP}} & s_{23}\ c_{13} \\ 
s_{12}\ s_{23}-c_{12}\ s_{13}\ c_{23}\ e^{i\delta_{CP}} & -c_{12}\ s_{23}-c_{23}\ s_{12}\ s_{13}\ e^{i\delta_{CP}} & c_{13}\ c_{23}
\end{array}\right), \label{unitary matrix}
\ee 
where $ c_{ij}\equiv\cos \theta_{ij}$, $ s_{ij}\equiv\sin \theta_{ij}$ and (i, j=1, 2, 3). In this study, we ignore the CP violating phase and will assume that it is equal to zero. The time evolution of neutrino flavor states during propagation is given by the following relation:
\be
\vert\nu_{\alpha}(t)\rangle= A_{\alpha e}(t)\vert\nu_{e}\rangle+ A_{\alpha \mu}(t)\vert\nu_{\mu}\rangle+ A_{\alpha \tau}(t)\vert\nu_{\tau}\rangle, \label{neutrino state}
\ee
in which $A_{\alpha\beta}(t)\equiv U_{\alpha i}^{\ast}\ e^{-iE_{i}t/\hbar}\ U_{\beta i}$.
Neutrino oscillation is the probability of transition from a flavor to the other one during the neutrino propagation. The discrepancy between mass and flavor eigenstates and also the non-zero and non-degenerate neutrino mass give rise to the neutrino oscillation.

 The hypothesis of entanglement between different modes of a single particle has been proposed in literature \cite{single particle} and experimentally demonstrated \cite{experimental}.  
Blasone et al. have considered flavor oscillations in neutrino as an example of multi-mode entanglement of single-particle states \cite{blasone entanglement, blasone concurrence}. Furthermore, they have analytically quantified the entanglement between different flavor modes by using a couple of  measurable quantities defined in the quantum resource theory framework. For instance, they have calculated the concurrence as a measure to quantify the entanglement, via wave packet approach for the three flavor neutrino oscillation \cite{blasone concurrence}. Another intresting quantumness aspect of neutrino oscillation is coherence which was considered by Song, Xue-Ke, et al \cite{quantifying coherence}. In fact, in that study the authors have quantified the coherence for neutrino oscillation by the $ l_{1}$-norm of coherence defined in quantum resource theory and then they have compared it with the experimentally observed data. Actually the neutrino oscillation takes place provided that we have non-zero off-diagonal elements in density matrix which leads to non-zero $ l_{1}$-norm \cite{glashow}. On the other hand, it is possible to have the entanglement between different modes of three flavor neutrinos only when the neutrino oscillation occurs. Hence, it is natural to expect a relationship between the coherence and entanglement in neutrino oscillations. Therefore, in this paper we compare the $ l_{1}$-norm as a coherence measure for three flavor neutrino oscillation with  the concurrence as a measure of entanglement computed by Blasone et al. \cite{blasone concurrence}. Of course, we may also name several other references that have
investigated quantum correlations in neutrino oscillation in addition to the
ones mentioned above such as Refs. \cite{quantum correlation alok, quantum information, ming, cp}. The quantumness
of neutrino oscillation can also be revealed by the Leggett-Garg inequality \cite{lgi violation in neutrino oscillation, lgi inequality in neutrino oscillation}.
In addition,  in order to have a complete and thorough description of neutrino oscillation, we should use the wave packet approach, since the production, propagation and detection of neutrino have to be considered as localized processes \cite{remaining coherent} and this localization is very well fulfilled using the wave packet approach in neutrino oscillation \cite{wave packet,paradoxes,dr.ettefaghi2015,dr.ettefaghi2020}. In this approach, one sees that a propagation length larger than the coherence length will lead the neutrino oscillation to diminish. Moreover, the coherence length is proportional to the wave packet width,  $\sigma_{x}$. Thus, in the following, we will investigate the variations of $ l_{1}$-norm in terms of $\sigma_{x}$.

In the next section, we will introduce the $l_{1}$-norm as a measure to quantify coherence in neutrino oscillations that is used by Xue-ke Song et al \cite{quantifying coherence}. We will also briefly introduce the concurrence for neutrino oscillation as computed by Blasone et al \cite{blasone concurrence}. In section 3, we will compute the $l_{1}$-norm for neutrino oscillation in three flavor framework via wave packet approach in order to study its variations by changing wave packet width. Finally in section 4, we will discuss our results and make the conclusion.

\section{Resource theory for coherence and entanglement in neutrino oscillation}

In this section, we will introduce a measure for coherence as well as entanglement according to quantum resource theories. We will then study those measures in the context of neutrino oscillation. 
 
Quantum coherence is one of the quantum correlations which can be considered as a quantum resource theory. Quantum resource theories provide several measures for coherence \cite{plenio2014, plenio2017}. Among them, there is the $ l_{1}$-norm which can be written as :
\be
{\cal C}(\rho)=\sum_{i\neq j}\vert\rho_{ij}\vert\geq 0 \label{coherence measure}
\ee
that is equal to the summation over the absolute values of all the off-diagonal elements $\rho_{ij}$  of a corresponding density matrix $ \rho $. The maximum possible value for ${\cal C}(\rho) $ is bounded by $ {\cal C}_{max}=d-1 $, where d is the dimension of the density matrix $ \rho $.

As has been mentioned, neutrino oscillation is one of the interesting quantum mechanical phenomena the quantumness of which can be determined with measures defined in quantum resource theories. Several measurements and analyses has been carried out on neutrino oscillation parameters. Recently the $l_{1}$-norm has been used for quantifying the quantumness of experimentally observed neutrino oscillation at Daya Bay, KamLAND and MINOS \cite{quantifying coherence}. For a neutrino  whose initial flavor in source is $\alpha $, the related density matrix in three generation framework can be written as:
\be
\rho_{\alpha}(t)=
\left(\begin{array}{ccc}
\vert A_{\alpha e}(t)\vert ^{2} & A_{\alpha e}(t)A_{\alpha \mu}^{\ast}(t) & A_{\alpha e}(t)A_{\alpha\tau}^{\ast}(t) \\ 
A_{\alpha e}^{\ast}(t)A_{\alpha\mu}(t) & \vert A_{\alpha\mu}(t)\vert^{2} & A_{\alpha\mu}(t)A_{\alpha\tau}^{\ast}(t)\\ 
A_{\alpha e}^{\ast}(t)A_{\alpha\tau}(t) & A_{\alpha\mu}^{\ast}(t)A_{\alpha\tau}(t) & \vert A_{\alpha\tau}(t)\vert^{2}
\end{array}\right).\label{coherence density} 
\ee
Since the dimension of the density matrix in this case is 3, $ {\cal C}_{max}=2$.
The $l_{1}$-norm  for this density matrix can be expressed in terms of the oscillation probabilities between different flavor modes as follows:
\be
{\cal C}_{\alpha}= 2(\sqrt{P_{\alpha e}P_{\alpha\mu}}+ \sqrt{P_{\alpha e}P_{\alpha\tau}}+ \sqrt{P_{\alpha\mu}P_{\alpha\tau}}). \label{coherence in terms of probabailities}
\ee

Entanglement is the other quantum correlation which is studied in the context of neutrino oscillation \cite{blasone entanglement, blasone concurrence}. As for coherence, there are several measures for entanglement in resource theory framework \cite{plenio1996, entanglement review, quantum correlations1, quantum correlations2}. Concurrence is one of these measures. For a state with a density matrix $\rho $, the concurrence measure is defined as \cite{quantum correlations1, quantum correlations2}:
\be
 C(\rho)=max(\lambda_{1}-\lambda_{2}-\lambda_{3}-\lambda_{4}, 0), \label{rho}
\ee
where $\lambda_{i}$ s are the square roots of the four eigenvalues of the non-Hermitian matrix  $\rho \tilde{\rho} $ in decreasing order with
\be
\tilde{\rho}=(\sigma_{y}\otimes\sigma_{y})\rho^{\ast}(\sigma_{y}\otimes\sigma_{y}).\label{rho tilde}
\ee 
Accordingly, if we consider a general two qubit pure state as
\begin{equation}
\vert\phi\rangle=\alpha_{00}\vert 0,0\rangle+ \alpha_{01} \vert 0,1\rangle+\alpha_{10}\vert 1,0\rangle+\alpha_{11}\vert 1,1\rangle, \label{pure state}
\end{equation} 
the concurrence is given by
\begin{equation}
{\cal C}(\rho)=2|\alpha_{00}\alpha_{11}-\alpha_{01}\alpha_{10}|.
\end{equation}

In three flavor scenario, the neutrino state can be considered as a three-qubit state:  $ \vert\nu_{e}\rangle \equiv \vert1\rangle_{\nu_{e}} \vert0\rangle_{\nu_{\mu}} \vert0\rangle_{\nu_{\tau}},\  \vert\nu_{\mu}\rangle\equiv\vert0\rangle_{\nu_{e}} \vert1\rangle_{\nu_{\mu}} \vert0\rangle_{\nu_{\tau}},\ \vert\nu_{\tau}\rangle\equiv\vert0\rangle_{\nu_{e}} \vert0\rangle_{\nu_{\mu}} \vert1\rangle_{\nu_{\tau}} $, in which the numbers 0 and 1 stand for the absence and presence of the corresponding flavors. Thus, the time-evolved of a neutrino state, i.e. relation (\ref{neutrino state}), can be considered as the  entanglement between different flavor modes in a single-particle as follows:
\be
\vert\nu_{\alpha}(t)\rangle= A_{\alpha e}(t)\vert1\rangle_{\nu_{e}} \vert0\rangle_{\nu_{\mu}} \vert0\rangle_{\nu_{\tau}} + A_{\alpha \mu}(t)\vert0\rangle_{\nu_{e}} \vert1\rangle_{\nu_{\mu}} \vert0\rangle_{\nu_{\tau}} +  A_{\alpha \tau}(t)\vert0\rangle_{\nu_{e}} \vert0\rangle_{\nu_{\mu}} \vert1\rangle_{\nu_{\tau}}.\label{entanglement}
\ee

In fact, the density matrix that corresponds to the above state, is an $8\times 8$ matrix which reduces to equation (\ref{coherence density}) after omitting the zero rows and columns. We should trace out one flavor  in equation (\ref{entanglement}), in order to obtain the desired concurrence between the other two flavor modes in the case of three flavor oscillation. In this way, we obtain a state similar to the state given by Eq. (\ref{pure state}) with $\alpha_{11}=0$ and consequently, the corresponding concurrence is due to the off-diagonal element of the reduced density matrix.
Thus we end up with three concurrences. Therefore, if the initial flavor is $\alpha $, the concurrence between flavors $\beta$ and $\gamma$ is obtained as follows:  
\be
C_{\beta \gamma}^{\alpha}=2\vert\sqrt{P_{\alpha\beta}P_{\alpha\gamma}}\vert. \label{concurrence}
\ee
The behavior of these quantities is investigated in terms of distance \cite{blasone concurrence}. However, we should note that the sum of three possible concurrences is equal to the $l_{1}$-norm, i.e. equation (\ref{coherence in terms of probabailities}), which quantified the quantum coherence. 
In fact the relation between quantum coherence and entanglement is the subject study of several articles such as Ref. \cite{plenio2017} and also many references mentioned in it such as Refs. \cite{relation between coherence and entanglement, coherence concurrence}. There it was shown
that any degree of coherence in the initial state of a
quantum system S can be converted to an entanglement
between S and the incoherent ancilla A by some incoherent operation. However, the above relation is different. In fact, a non-coherent and non-entangled state at the initial time (initial neutrino state in flavor basis) is evolved by time evolution operator to a coherent and entangled state.  Therefore, the origin of the entanglement between the flavor modes in a single-particle defined by Refs. \cite{blasone entanglement, blasone concurrence}, is the same as that of the quantum coherence.

  {Furthermore, we should note that the $l_1$-norm and the concurrences are written in terms of oscillation probabilities. Thus, this correspondence always holds regardless of the CP violation phase value. Meanwhile, the behavior of $l_1$-norm in terms of CP violation phase is investigated in Ref. \cite{cp}.

\section{Calculation of $l_{1}$-norm for neutrino oscillation via wave packet approach}

In this section, we will calculate the $l_{1}$-norm for neutrino oscillation in three flavor framework via the wave packet approach. We intend to investigate the relation between the quantum coherence and the wave packet width, $\sigma_{x}$. The transition amplitudes in the wave packet approach will take the following form \cite{wave packet}:
\be
A_{\alpha\beta}(L, T)=\sqrt{\dfrac{2\sigma_{xD}\sigma_{xP}}{\sigma_{x}^{2}}}\ \sum_{a}U_{\alpha a}^{\ast}U_{\beta a}\ \exp[-iE_{a}T+iP_{a}L-\dfrac{(L-v_{a}T)^{2}}{4\sigma_{x}^{2}}],\label{wave packet amplitudes}
\ee 
in which
\be
\sigma_{x}^{2}\equiv\sigma_{xD}^{2}+\sigma_{xP}^{2}.\label{wave packet width}
\ee 
In this equation, $\sigma_{xP}$ and $ \sigma_{xD}$ correspond to the produced and detected neutrino wave packet widths. To calculate the oscillation probability, the following approximations can be considered in equation (\ref{wave packet amplitudes}):
\be
P_{a}\cong\ E-(1-\zeta)\dfrac{m_{a}^{2}}{2E},\  E_{a}\cong E+ \zeta \dfrac{m_{a}^{2}}{2E}, \ v_{a}\cong 1-\dfrac{m_{a}^{2}}{2E},\label{wave packet probabilities}
\ee 
where $ \zeta=\dfrac{1}{2}(1-\dfrac{m_{\mu}^{2}}{m_{\pi}^{2}}) =0.2 $ with $m_{\mu}$ and $m_{\pi}$ denoting the muon and pion rest masses, respectively. Also $ v_{a} $ is the wave packet group velocity. The existing parameters in equation (\ref{wave packet amplitudes}),  such as the mixing angles $\theta_{ij}$, squared mass difference $ \Delta m_{ab}^{2}= m_{a}^{2}- m_{b}^{2}$ and the parameter E will be fixed by their experimental values. After parameterizing the three possible mass squared differences in terms of the two independent parameters, $  \delta m^{2}$ and $\Delta m^{2}$:
\be
\Delta m_{12}^{2}= \delta m^{2},\quad \Delta m_{31}^{2}=\Delta m^{2}+ \dfrac{\delta m^{2}}{2},\quad \Delta m_{32}^{2}=\Delta m^{2}- \dfrac{\delta m^{2}}{2}, \label{parameterizing}
\ee
we assume the following experimental values for the related parameters:
\bea
&&\sin ^{2} \theta_{12}= 0.314,\quad \sin ^{2} \theta_{13}= 0.8\times 10^{-2},\quad \sin ^{2} \theta_{23}= 0.45,\nn\\
&&\Delta m^{2}= 2.6\times 10^{-3}\ eV^{2},\quad \delta m^{2}= 7.92\times 10^{-5} eV^{2},\quad E= 10 GeV. \label{experimental values}
\eea

Applying the mentioned approximations and above assumptions, one can write the transition probabilities in terms of distance and time. Since the propagation time cannot be measured, we must integrate over time T. Thus, the transition probabilities will be given in terms of distance:
\be
P_{\alpha\beta}(L)=\sum_{a, b} U_{\alpha a}^{\ast} U_{\beta a} U_{\alpha b} U_{\beta b}^{\ast}\ \exp [-2\pi i\dfrac{L}{L_{ab}^{osc}}-(\dfrac{L}{L_{ab}^{coh}})^{2}]F_{ab}, \label{probability}
\ee
in which:
\be
L_{ab}^{osc}=\dfrac{4\pi E}{\Delta m_{ab}^{2}},\ L_{ab}^{coh}=\dfrac{4\sqrt{2}\sigma_{x}E^{2}}{\vert \Delta m_{ab}^{2}\vert}\label{coherence length}
\ee
and:
\be
F_{ab}=\exp[-2\pi^{2}(1-\zeta)^{2} (\dfrac{\sigma_{x}}{L_{ab}^{osc}})^{2}]. \label{f}
\ee
In this manner, we have two more factors in comparison to the plane wave approach. The first factor is the Guassian factor which describes the coherence condition. According to this condition, the oscillation can only take place if the overlapping of the mass eigenstates is not destroyed; which means $L < L^{coh}$. In fact, as it can be seen from the equation (\ref{coherence length}), the wave packet width $\sigma_{x}$ plays the main role, since the coherence length is proportional to $\sigma_{x}$. That is to say, the larger the wave packet width, the larger the coherence length and consequently we can see the oscillation at larger distances. The second factor is $F_{ab}$ which emphasizes that for seeing the oscillation we must have $\sigma_{x}\ll L_{ab}^{coh}$. Here we study the relation between the $l_{1}$-norm as a measure of coherence and $\sigma_{x}$.

In Fig.1, we plot the $l_{1}$-norm for three different values of  $ \sigma_{x} $ in terms of distance, when the initial neutrino is electron. We observe that the amount of coherence, increases with $\sigma_{x}$. The expected result is achieved because, when $\sigma_{x}$ increases, the overlapping of the wave packets survives for larger distances and therefore the $l_{1}$-norm increases.

\begin{figure}[h!]
\begin{center}
\includegraphics{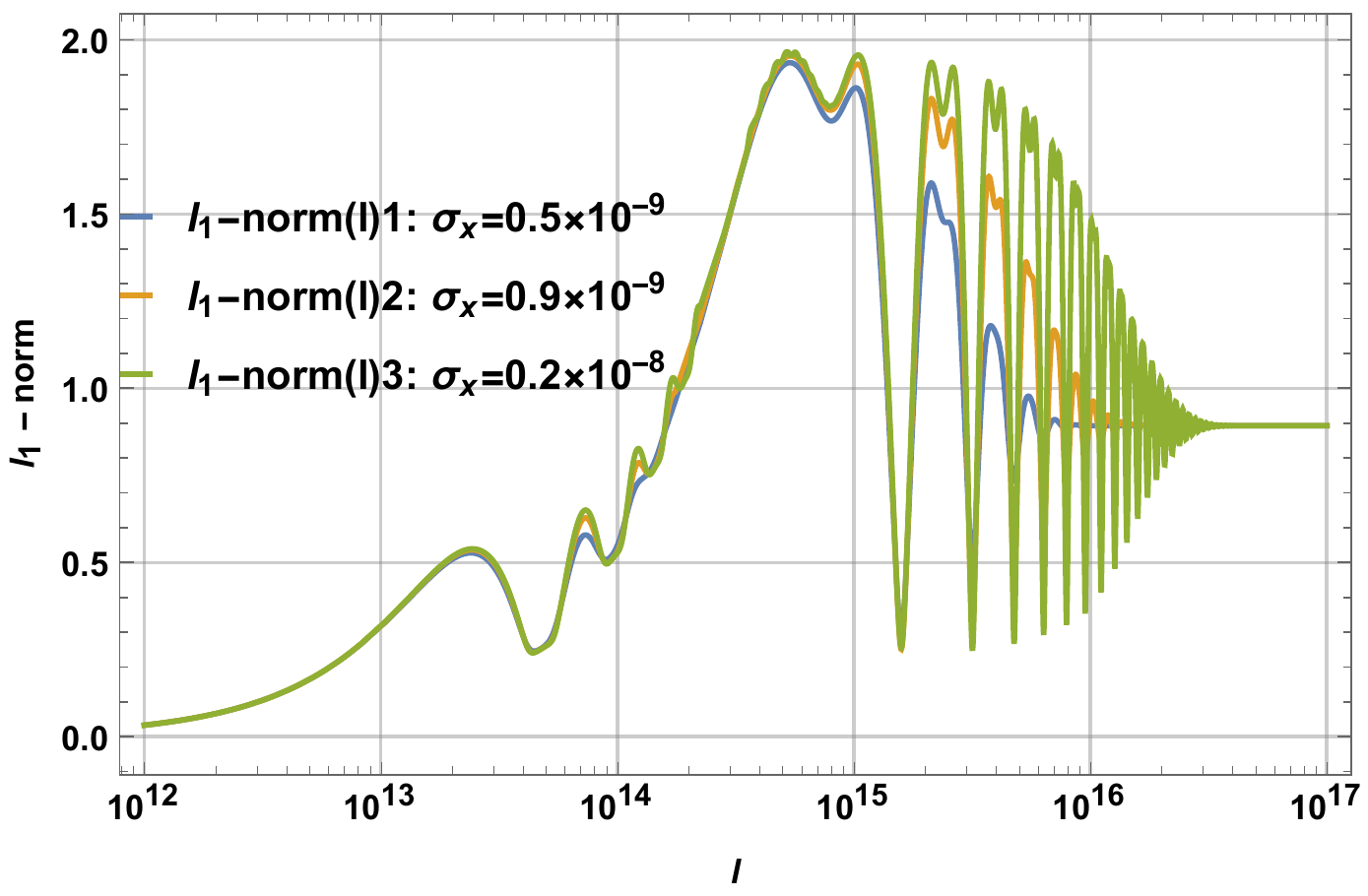}
\caption{Variations in $l_{1}$-norm as a function of distance l in kilometers, for three different wave packet widths, $ \sigma_{x}$. } \label{1}
\end{center}
\end{figure}	

\section{Conclusion}

	Neutrino oscillation is an interesting and exciting quantum phenomenon which provides an excellent opportunity for studying quantum effects in large distances. In particular, the coherence  and entanglement as two of the most famous quantum correlations were investigated for this system in Ref. \cite{ quantifying coherence} and Refs. \cite{blasone entanglement, blasone concurrence}, repectively. In this paper, we have shown that the $l_{1}$-norm as the coherence measure and the concurrence as the entanglement measure  are related to each other in the quantum resource theory framework. Actually, the coherence measure equals sum of the three possible concurrences between different neutrino flavor modes which were obtained by Blasone et al. \cite{blasone concurrence}.  This relation shows a correspondence between quantum coherence and entanglement for neutrino oscillation. This correspondence is different from what were discussed in Refs. \cite{plenio2017,relation between coherence and entanglement,coherence concurrence}. There, a coherent state is entangled with a non-coherent state through a non-coherent operator, thus creating a correspondence between the coherence and entanglement  quantum resource theories. In the case of neutrino oscillation,  the initial state which is a non-coherent and non-entangled state in the flavor basis, becomes a coherent and entangled state, after the time evolution. Therefore, we see another correspondence between the coherence and entanglement quantum resource theories. This relation holds independent of neutrino oscillation parameters such as CP violation phase. This seems to be natural, since the entanglement between different flavor modes is due to the existence of coherence.
	
	 Moreover, using the wave packet approach, we have calculated the $l_{1}$-norm for three flavor neutrino oscillation. The variations of this coherence measure has been plotted in terms of distance with changing the wave packet width, $ \sigma_{x} $ (please see Fig.\ref{1}). It is observed that the amount of coherence increases by $ \sigma_{x} $. We expect this result, because the overlapping of the mass eigenstates increases by $\sigma_{x}$ and therefore there will be more coherence. 

	Neutrinos interact only through weak interactions. Therefore, their quantum coherence can be preserved over astrophysical length scales. This is an advantage for neutrinos in comparison to the other particles widely utilized for quantum-information processing. There exist a number of works in which neutrinos are purposed to be used for communication, for instance, see Refs. \cite{neutrino communication1, neutrino communication2, neutrino communication3}.  For future applications of neutrinos in quantum information, it will be significant to have an accurate description of the appropriate quantum resource theory introducing a measure that quantifies the quantum correlations of neutrino. Furthermore, the probabilities of neutrino oscillation as well as all the other quantum correlations are modified when neutrinos propagate through large distances in matter with high density. Hence, studying the quantum resource theories for neutrino correlations in matter will steer us to improve the performance of neutrino quantum-information processing.

\end{document}